\pdfoutput=1

\documentclass[12pt]{article}
\usepackage{amsmath,epsf,amssymb,latexsym,cite,shadow,eucal,theorem,enumerate}
\usepackage[matrix,arrow,frame,import,curve,color]{xy}
\usepackage{tikz}
\usetikzlibrary{calc}
\usetikzlibrary{arrows}
\usetikzlibrary{decorations.pathreplacing,decorations.markings}

\usepackage{graphicx}

\newtheorem{prop}{Proposition}

{\theorembodyfont{\rmfamily}}

\newif\iffigs\figstrue

\DeclareFontFamily{U}{rsf}{}
\DeclareFontShape{U}{rsf}{m}{n}{
  <5> <6> rsfs5 <7> <8> <9> rsfs7 <10-> rsfs10}{}
\DeclareMathAlphabet\Scr{U}{rsf}{m}{n}

\def\O{\Scr{O}}
\def\C{{\mathbb C}}
\def\P{{\mathbb P}}

\def\Q{{\mathbb Q}}

\def\Z{{\mathbb Z}}

\def\sHom{\operatorname{\Scr{H}\!\!\textit{om}}}

\def\Ext{\operatorname{Ext}}

\def\Spec{\operatorname{Spec}}

\def\Sl{\operatorname{SL}}

\def\CY{Calabi--Yau}
\def\LG{Landau--Ginzburg}

\def\cN{{\Scr N}}

\def\cA{{\Scr A}}

\def\cI{{\Scr I}}

\def\cT{{\Scr T}}

\def\cE{{\Scr E}}

\def\DC{\mathbf{D}}

\def\QED{$\quad\blacksquare$}
\def\ff#1#2{{\textstyle\frac{#1}{#2}}}

\def\eqn#1#2{\begin{equation}#2
  \ifx{#1}{}\else\label{#1}\fi\end{equation}}

\def\arrj#1#2#3{
  \coordinate (m) at ($ (#1)!.5!(#2) $);
  \draw[->] (#1) -- (m);
  \draw (m) -- (#2);
  \node[inner sep=0pt,label=above:$\scriptstyle #3$] at (m) {}
}

\def\arrja#1#2#3{
  \coordinate (m) at ($ (#1)!.5!(#2) $);
  \draw[->] (#1) -- (m);
  \draw (m) -- (#2);
  \node[inner sep=0pt,label=right:$\scriptstyle #3$] at (m) {}
}

\def\arrjb#1#2#3{
  \coordinate (m) at ($ (#1)!.5!(#2) $);
  \draw[->] (#1) -- (m);
  \draw (m) -- (#2);
  \node[inner sep=0pt,label=left:$\scriptstyle #3$] at (m) {}
}

\def\arrjd#1#2#3{
  \coordinate (m) at ($ (#1)!.5!(#2) $);
  \draw[->] (#1) -- (m);
  \draw (m) -- (#2);
  \node[inner sep=0pt,label=below:$\scriptstyle #3$] at (m) {}
}

\tikzset{
  on each segment/.style={
    decorate,
    decoration={
      show path construction,
      moveto code={},
      lineto code={
        \path [#1]
        (\tikzinputsegmentfirst) -- (\tikzinputsegmentlast);
      },
      curveto code={
        \path [#1] (\tikzinputsegmentfirst)
        .. controls
        (\tikzinputsegmentsupporta) and (\tikzinputsegmentsupportb)
        ..
        (\tikzinputsegmentlast);
      },
      closepath code={
        \path [#1]
        (\tikzinputsegmentfirst) -- (\tikzinputsegmentlast);
      },
    },
  },
  mid arrow/.style={postaction={decorate,decoration={
        markings,
        mark=at position .47 with {\arrowreversed[#1]{stealth}}
      }}},
}

\begin{document}

\begin{titlepage}
\begin{flushright}
April 2014
\end{flushright}
\vspace{.5cm}
\begin{center}
\baselineskip=16pt
{\fontfamily{ptm}\selectfont\bfseries\huge
Rational Curves and (0,2)-Deformations\\[20mm]}
{\bf\large  Paul S.~Aspinwall and Benjamin Gaines
 } \\[7mm]

{\small

Department of Mathematics\\ Duke University, 
 Durham, NC 27708-0320 \\ \vspace{6pt}

 }

\end{center}

\begin{center}
{\bf Abstract}
\end{center}
We compare the count of $(0,2)$-deformation moduli fields for
$N=(2,2)$ conformal field theories on orbifolds and sigma-models on resolutions
of the orbifold. The latter involves counting deformations of the
tangent sheaf. We see there is generally a discrepancy which is
expected to be explained by worldsheet instanton corrections coming from
rational curves in the orbifold resolution. We analyze the rational
curves on the resolution to determine such corrections and discover
that irreducible toric rational curves account for some, but not all,
of the discrepancy. In particular, this proves that there must be
worldsheet instanton corrections beyond those from smooth isolated
rational curves.  \vfil\noindent

\end{titlepage}

\vfil\break


\section{Introduction}    \label{s:intro}

Using comparisons between conformal field theory and geometry to
predict facts about rational curves has been famously successful in
the context of mirror symmetry \cite{CDGP:}. In this paper we will
make another such comparison which should be quite distinct as it
lives outside the world of mirror symmetry and topological
field theory.

Here we are concerned with (0,2)-superconformal field theories
obtained from a non-linear $\sigma$-model associated to a Calabi-Yau
threefold $X$ equipped with a rank 3 holomorphic vector bundle $V\to
X$.  Let $\cT$ be the tangent bundle of $X$.  We consider the case
where the (0,2) superconformal field theory is a first order
deformation of a (2,2) theory. Geometrically this means that $V$ is a
first order deformation of $\cT$.

We will consider the case where $X$ is a projective resolution of an orbifold
$\C^3/G$ and compare the number of deformations of the conformal field
theory to the number of deformations of $\cT$. Any discrepancy should
be due to worldsheet instantons which probes the structure of
rational curves in $X$.

The subject of worldsheet instanton corrections in non-linear
$\sigma$-models has a fascinating history. They were first analyzed in
the sequence of papers \cite{Wen:1985jz,DSWW:}. Here it
was realized that typically they could produce corrections to the
superpotential resulting in one-point functions that would destabilize
the vacuum. That is, the combination $V\to X$ could not be used for
string compactifications.

The special case, where $V$ is the tangent bundle $\cT$, leads to
(2,2) worldsheet supersymmetry. It is well-known that these give
perfectly good string compactifications. Here the instantons do not destroy
the vacuum but do correct certain three-point functions. Analysis of such
corrections by using mirror symmetry to compare geometry and exact
conformal field theory \cite{CDGP:,AM:rat} led to the present day
flourishing of Gromov--Witten theory (see, for example,
\cite{CK:mbook}).

Meanwhile it was realized that some other (0,2)-models might actually
be okay too. If the bundle $V$ is restricted to a rational curve
$C\subset X$ then it will always split
\begin{equation}
  \left.V\right|_C = \O(a)\oplus\O(b)\oplus\O(c),  \label{eq:split}
\end{equation}
where $a+b+c=0$, since $c_1(V)=0$. In \cite{Dist:res,Distler:1987ee}
it was emphasized that, in the case of isolated (to first order)
curves, instantons could only produce those dangerous one-point
functions if $V$ split ``trivially'', i.e., $a=b=c=0$ . So if $V$
splits nontrivially on {\em every\/} rational curve, the model is
good. This explains why (2,2)-models are good, since $V\cong\cT$ and
adjunction implies $a\geq2$. Unfortunately, as we review in appendix,
$a=b=c=0$ is the {\em generic\/} case for a bundle over $\P^1$. Thus,
unless there are global obstructions in $X$ to prevent such a
splitting, a generic deformation of $V$ away from a (2,2)-model would
seem to be bad.  So generic $V\to X$ still tend to give invalid models.

There is another way of killing the instanton corrections. It may be
that each rational curve gives a nonzero contribution to one-point
functions, but these contributions magically cancel when all the
rational curves are added up. Such a miracle indeed happens for the
quintic threefold \cite{Silverstein:1995re} as was explained further
in \cite{Beasley:2003fx}. The generic quintic has 2875 lines and over
each of these lines a generic deformation of $\cT$ will split
trivially. However, such a deformation of $\cT$ always has zero net
instanton corrections to one-point functions since the contribution
from these lines cancel. Similarly all higher degree rational curves
cancel too.

Because of this, (0,2)-models sometimes seem to be plagued by a
contradiction: even though a generic model should be killed by
instanton effects, anything you can actually write down (and, in
particular, know the exact conformal field theory) probably isn't
generic and is perfectly fine! More precisely, the analysis of
\cite{Beasley:2003fx} suggests that anything that can be written as a
gauged linear $\sigma$-model evades instanton effects.

It was recently observed in \cite{AP:elusive,me:02mc} that resolutions of
orbifolds can see instanton effects. Here we really can compare
conformal field theory and geometry to get nontrivial statements. 
It was shown in \cite{me:02mc} that while the count of deformations of
the conformal field theory makes no reference to a choice of
resolution, the number of deformations of the tangent sheaf does
depend on this choice. This difference must be due to
instantons. 

Interestingly, one favoured choice of resolution, the Hilbert scheme,
always agrees with conformal field theory \cite{BenG:Hilb}. Instantons
are therefore associated with deviations of a resolution from the
Hilbert scheme. We will use this approach to systematically look for
instanton effects below.

In section \ref{s:count} we will review how to count the number of
(0,2)-deformations in an orbifold conformal field theory and how to
count deformations of the tangent bundle of a noncompact toric \CY\
threefold. Simple diagrammatical methods using quivers from the toric
data from \cite{me:02mc} make the analysis straight-forward even
though we need to use fairly large groups for $G$.

In section \ref{s:inst} we review the necessary material for
worldsheet instantons and how splitting types of bundles is
central. In section \ref{s:split} we analyze the geometry of splitting
and see exactly how to hunt for instantons in the case the rational
curves are smooth and toric.  While we find some instantons, we do not
find enough and in section \ref{s:oth} we briefly discuss other
possible sources such as reducible curves and multiple covers.  We
should emphasize that since we are doing (0,2)-deformations, we do
{\em not\/} simply reproduce Gromov--Witten theory in such cases.


\section{Counting Deformations}   \label{s:count}

In this section we review the results of \cite{me:02mc} where we see
the discrepancy between the orbifold conformal field theory
computation and the geometric computation of the number of
deformations of the tangent bundle. 

Let $G\subset\Sl(3)$ be a finite Abelian group. We are then
considering the orbifold $\C^3/G$ and we will assume that $G$ has an
{\em isolated\/} fixed point at the origin. We denote the resolution
of $\C^3/G$ by $X$ or $X_\Sigma$.

\subsection{Orbifold Conformal Field Theory}  \label{s:CFT}

Let $g\in G$ and diagonalize the action on $\C^3$ as
\begin{equation}
(z_1,z_2,z_3)\mapsto (e^{2\pi i\nu_1}z_1,e^{2\pi i\nu_2}z_2,e^{2\pi
i\nu_3}z_3),
\end{equation}
where $0\leq\nu_i<1$. Define $\tilde\nu_i=\nu_i-\ff12-m$, where
$m\in\Z$ such that $-1<\tilde\nu_i\leq0$.

We may enumerate the number of massless fields corresponding to
deformations as follows paraphrasing \cite{meMP:singlets,me:02mc}. We
get contributions from each $g$-twisted sector when $g$ satisfies
\begin{equation}
  \sum_i\nu_i=1.  \label{eq:junior}
\end{equation}
Note that in this case, at most one of the $\nu_i$'s can be greater
than $\ff12$.  For each $g$-twisted sector satisfying (\ref{eq:junior}) the
contribution is given by the coefficient of $q^0z^0$ in the power
series expansion of the partition function
\begin{equation}
Z = q^az^b\prod_{i=1}^3 \frac{\left(1+z^{-1}q^{1+\tilde\nu_i}\right)
 \left(1+zq^{-\tilde\nu_i}\right)}
{\left(1-q^{\nu_i}\right)\left(1-q^{1-\nu_i}\right)},
\end{equation}
where
\begin{equation}
(a,b) = \begin{cases} (-\ff12,-1) &\hbox{if all $\nu_i\leq\ff12$}\\
(-\nu_i,0)&\hbox{if $\nu_i>\ff12$}\end{cases}
\end{equation}
Adding the contributions of all such $g$'s together we get the total
number of first order deformations. Some of these are deformations
preserving the (2,2) supersymmetry (counted by the number of $g$'s
satisfying (\ref{eq:junior})) and correspond to blow-up K\"ahler
deformations. The remainder should be deformations of the tangent
bundle.

A key point to emphasize here is that the conformal field theory count 
makes no reference to a resolution of the singularity and thus,
obviously, cannot depend on the choice of resolution.

It should be noted that the number of (0,2)-deformations is not an
invariant over the moduli space and can increase on lower-dimensional
strata \cite{Berglund:1990rk}. This can most
easily be seen in \LG\ models and was analyzed in detail in
\cite{meMP:singlets}. This effect does not seem to be present in
resolutions of $\C^3/G$. The Hilbert scheme count agrees perfectly
with the conformal field theory count \cite{BenG:Hilb}. It is possible
that the orbifold conformal field theory count has ``extra'' moduli
which disappear when the blow-up is turned on. Then the Hilbert scheme
would need to have instanton corrections to lower its count by the
same margin. This coincidence seems very unlikely and we prove that
there are no instantons in the form of toric curves later in this
paper. We will therefore ignore this possibility. It certainly does
not help resolve the issue of lack of instantons in this paper --- it
would make it worse.

\subsection{Geometry}   \label{ss:geom}

Now we review how to compute the number of deformations of the tangent
bundle of a toric variety. To fix notation, we need to quickly review
the basic construction of toric geometry. We refer to \cite{Cox:} for
all the details. Begin with
\begin{equation}
\xymatrix@1@C=20mm{
0\ar[r]&\mathsf{M}\ar[r]^-{\cA}&\Z^{\oplus n}\ar[r]^-\Phi&\mathsf{D}\ar[r]&0,
} \label{eq:toric}
\end{equation}
where $\mathsf{M}$ is a lattice of rank $d$. $\cA$ is an $n\times d$
matrix. The rows of $\cA$ give the coordinates of a set of $n$ points
in the lattice $\mathsf{N}$ dual to $\mathsf{M}$. We will use the same
symbol $\cA$ to denote this point set. Each point is associated to a
homogeneous coordinate $x_i$, $i=1,\ldots,n$. We then have a
homogeneous coordinate ring $S=\C[x_1,\ldots,x_n]$.

$\mathsf{D}$ is an abelian group which induces an action on the
homogeneous coordinates as follows. Let $r=n-d$ denote the rank of
$\mathsf{D}$. Then $\mathsf{D}\otimes_\Z\C^*=(\C^*)^r$ acts on the
homogeneous coordinates with charges given by the matrix $\Phi$. If
$\mathsf{D}$ has a torsion part $H$, then, in addition to $(\C^*)^r$,
we have a finite group action of $H$ on the homogeneous coordinates
the charges of which are given by the kernel of the matrix $\cA^t$ acting
on $(\Q/\Z)^{\oplus n}$.

The fan $\Sigma$ is defined as a fan over a simplicial
complex with vertices $\cA$. This combinatorial information defines an
ideal $B\subset S$ as explained in \cite{Cox:}. The toric variety
$X_\Sigma$ is then given by the quotient
\begin{equation}
  X_\Sigma = \frac{\Spec S - V(B)}{(\C^*)^r\times H}.
\end{equation}

The homogeneous coordinate ring $S=\C[x_1,\ldots,x_n]$ is multigraded
by the free module $\mathsf{D}$:
\begin{equation}
  S = \bigoplus_{\mathbf{d}\in\mathsf{D}} S_{\mathbf{d}}.
\end{equation}
Let $S(\mathbf{q})$ be the free $S$-module with grades shifted so that
$S(\mathbf{q})_{\mathbf{d}}=S_{\mathbf{q}+\mathbf{d}}$ as usual. Line
bundles or invertible sheaves $\O(\mathbf{q})$ on $X_\Sigma$ are then
associated to modules $S(\mathbf{q})$ for any
$\mathbf{q}\in\mathsf{D}$.

In the case $X$ is a \CY\ variety, all the points $\cA$ lie in a
hyperplane in $\mathsf{N}$ . The orbifold singularity $\C^3/G$ can
always be completely resolved crepantly by taking $\cA$ to contain all
the points of $\mathsf{N}$ in its interior hull and taking $\Sigma$ to
be a fan over a simplicial complex including all the new points of
$\cA$. Let $\Delta$ denote this simplicial complex. Each triangle in
$\Delta$ has area $\ff12$, $\mathsf{D}$ is torsion-free and $X_\Sigma$
is smooth. Note that the condition that $G$ has an isolated fixed
point means that points in $\cA$ are either vertices of the triangle
$\Delta$ or properly in the interior.

The supersymmetric conformal field theory requires a K\"ahler form on
$X$ and so we insist $X$ is projective. This requires the
triangulation of $\Delta$ to be ``regular'' or ``coherent'' in the
sense of \cite{GKZ:book}.

\begin{figure}
\begin{center}
\begin{tikzpicture}[x=0.3mm,y=0.3mm]
  \path[shape=circle,inner sep=1pt,minimum size=5mm,every node/.style={draw}]
     (0,0) node(a0) {$\alpha$}
     (50,50) node(a1) {$j$}
     (0,50) node(a2) {$i$}
     (-30,100) node(a4) {$\beta$};
  \arrj{a2}{a1}{m};
  \draw (a4) -- (a1) -- (a0) -- (a2) -- (a4);
  \draw(17,44) node {$\scriptstyle C$};
  \draw(-15,-12) node {$\scriptstyle (0,-1)$};
  \draw(-15,38) node {$\scriptstyle (0,0)$};
  \draw(65,38) node {$\scriptstyle (1,0)$};
  \draw(-45,84) node {$\scriptstyle (1-m,1)$};
\end{tikzpicture}
\end{center}
\caption{A toric curve $C=D_i\cap D_j$.} \label{fig:C}
\end{figure}
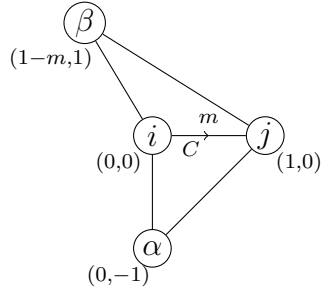

A {\em toric rational curve\/} is a rational curve in $X_\Sigma$ which
is invariant under the $(\C^*)^d$ torus action. Because of the
``orbit-cone correspondence'', such rational curves correspond to
edges of the graph giving the triangulation of $\Delta$. For the curve
to be compact, the edge must be in the interior of $\Delta$. Thus, the
edge is the edge of two triangles. By a choice of coordinates on
$\Delta$, we may always make the edge look like figure \ref{fig:C}
for some $m\in\Z$. Points $i,j,\alpha,\beta\in\cA$ correspond to divisors
which we denote $D_i$ etc. The rational
curve is $C=D_i\cap D_j$. This curve intersects each of the
divisors $D_\alpha$ and $D_\beta$ transversely at a point.  We will
review in section \ref{ss:rst} that the normal bundle
of this curve is then
\begin{equation}
  \cN = \O(-1-m)\oplus\O(-1+m),
\end{equation}
and we refer to it as a $(-1-m,-1+m)$-curve. For each curve
$C_{ij}=D_i\cap D_j$ we may assume $m_{ij}\geq0$ by reversing $i$ and
$j$ in figure \ref{fig:C} as necessary. If $m>0$ this fixes the
direction of the arrow in the figure --- it always points ``away''
from $\beta$. These arrows will be important when analyzing instanton
corrections later in the paper. When $m=0$ there will be no arrow and
we will draw a dotted line between $i$ and $j$. Note that the
coordinates $x_\alpha$ and $x_\beta$ associated with the corresponding
points in figure \ref{fig:C} can be used as the homogeneous
coordinates on $C_{ij}$.

We will explore how to count deformations of the tangent bundle in
section \ref{ss:defs} but, for now, we will just quote the result from
\cite{me:02mc}.  The number of blow-up modes plus framed deformations
of the tangent bundle is given by
\begin{equation}
  3n_{\textrm{int}} + \sum_{C_{ij}} m_{ij},  \label{eq:geomcount}
\end{equation}
where $n_{\textrm{int}}$ is the number of internal points in
$\Delta$. Each toric curve $C_{ij}$ is counted once in
(\ref{eq:geomcount}) where we order $i$ and $j$ so that $m_{ij}\geq0$.

\subsection{A $\Z_{11}$ Example}   \label{ss:Z11}

To illustrate the above computations consider $\C^3/G$ where
$G\cong\Z_{11}$ generated by $g$ with $(\nu_1,\nu_2,\nu_3)=
(\ff1{11},\ff2{11},\ff8{11})$. The elements of $G$ satisfying
(\ref{eq:junior}) are then $g,g^2,g^3,g^6,g^7$ which yield 14,4,5,7,9
moduli respectively. Thus the exact conformal field theory computation
dictates that there must be 39 (0,2) deformations (of which 5
are associated to blow-up modes giving (2,2) deformations).

\begin{figure}
\begin{center}
\begin{tikzpicture}[x=0.8mm,y=0.8mm]
  \path[shape=circle,inner sep=1pt,every node/.style={draw}]
     (0,0) node(a0) {1}
     (50,0) node(a1) {2}
     (25,43.3) node(a2) {3}
     (27.27,31.49) node(a3) {4}
     (29.54,19.68) node(a4) {5}
     (31.81,7.87) node(a5) {6}
     (13.64,15.75) node(a6) {7}
     (15.91,3.94) node(a7) {8};
  \draw[dashed] (a0) -- (a1) -- (a2) -- (a0);
  \draw[dashed] (a4) -- (a6) -- (a7) -- (a4);
  \arrja{a3}{a2}{5};
  \arrj{a3}{a1}{1};
  \arrj{a4}{a1}{1};
  \arrj{a5}{a1}{2};
  \arrja{a5}{a4}{2};
  \arrja{a4}{a3}{3};
  \arrj{a3}{a6}{1};
  \arrj{a5}{a7}{2};
  \arrjd{a7}{a1}{1};
  \arrj{a7}{a0}{3};
  \arrj{a6}{a0}{2};
  \arrj{a6}{a2}{1};
  \draw[every node/.style={draw}] (10,30) node {39};
\end{tikzpicture}
\qquad
\begin{tikzpicture}[x=0.8mm,y=0.8mm]
  \path[shape=circle,inner sep=1pt,every node/.style={draw}]
     (0,0) node(a0) {1}
     (50,0) node(a1) {2}
     (25,43.3) node(a2) {3}
     (27.27,31.49) node(a3) {4}
     (29.54,19.68) node(a4) {5}
     (31.81,7.87) node(a5) {6}
     (13.64,15.75) node(a6) {7}
     (15.91,3.94) node(a7) {8};
  \draw[dashed] (a0) -- (a1) -- (a2) -- (a0);
  \draw[dashed] (a5) -- (a6);
  \arrj{a6}{a0}{2};
  \arrj{a7}{a0}{3};
  \arrj{a6}{a2}{1};
  \arrja{a3}{a2}{5};
  \arrj{a3}{a6}{1};
  \arrja{a4}{a3}{3};
  \arrja{a5}{a4}{1};
  \arrja{a7}{a6}{1};
  \arrj{a3}{a1}{1};
  \arrj{a4}{a1}{1};
  \arrj{a5}{a1}{2};
  \arrj{a4}{a6}{1};
  \arrj{a5}{a7}{1};
  \arrjd{a7}{a1}{1};
  \draw[every node/.style={draw}] (10,30) node {39};
\end{tikzpicture}\\[7mm]

\begin{tikzpicture}[x=0.8mm,y=0.8mm]
  \path[shape=circle,inner sep=1pt,every node/.style={draw}]
     (0,0) node(a0) {1}
     (50,0) node(a1) {2}
     (25,43.3) node(a2) {3}
     (27.27,31.49) node(a3) {4}
     (29.54,19.68) node(a4) {5}
     (31.81,7.87) node(a5) {6}
     (13.64,15.75) node(a6) {7}
     (15.91,3.94) node(a7) {8};
  \draw[dashed] (a0) -- (a1) -- (a2) -- (a0);
  \draw[dashed] (a6) -- (a3) -- (a7);
  \arrja{a3}{a2}{5};
  \arrj{a3}{a1}{1};
  \arrj{a4}{a1}{1};
  \arrj{a5}{a1}{2};
  \arrja{a5}{a4}{2};
  \arrja{a4}{a3}{4};
  \arrj{a5}{a7}{2};
  \arrjd{a7}{a1}{1};
  \arrj{a7}{a0}{3};
  \arrj{a6}{a0}{2};
  \arrj{a6}{a2}{1};
  \arrj{a4}{a7}{1};
  \arrja{a6}{a7}{1};
  \draw[every node/.style={draw}] (10,30) node {41};
\end{tikzpicture}
\begin{tikzpicture}[x=0.8mm,y=0.8mm]
  \path[shape=circle,inner sep=1pt,every node/.style={draw}]
     (0,0) node(a0) {1}
     (50,0) node(a1) {2}
     (25,43.3) node(a2) {3}
     (27.27,31.49) node(a3) {4}
     (29.54,19.68) node(a4) {5}
     (31.81,7.87) node(a5) {6}
     (13.64,15.75) node(a6) {7}
     (15.91,3.94) node(a7) {8};
  \draw[dashed] (a4) -- (a0) -- (a1) -- (a2) -- (a0);
  \arrja{a3}{a2}{5};
  \arrj{a3}{a1}{1};
  \arrj{a4}{a1}{1};
  \arrj{a5}{a1}{2};
  \arrja{a5}{a4}{2};
  \arrja{a4}{a3}{3};
  \arrj{a5}{a7}{2};
  \arrjd{a7}{a1}{1};
  \arrj{a7}{a0}{4};
  \arrj{a6}{a0}{3};
  \arrj{a6}{a2}{1};
  \arrj{a7}{a4}{1};
  \arrj{a3}{a6}{1};
  \arrj{a6}{a4}{1};
  \draw[every node/.style={draw}] (10,30) node {43};
\end{tikzpicture}
\begin{tikzpicture}[x=0.8mm,y=0.8mm]
  \path[shape=circle,inner sep=1pt,every node/.style={draw}]
     (0,0) node(a0) {1}
     (50,0) node(a1) {2}
     (25,43.3) node(a2) {3}
     (27.27,31.49) node(a3) {4}
     (29.54,19.68) node(a4) {5}
     (31.81,7.87) node(a5) {6}
     (13.64,15.75) node(a6) {7}
     (15.91,3.94) node(a7) {8};
  \draw[dashed] (a7) -- (a2) -- (a1) -- (a0) -- (a2);
  \arrja{a3}{a2}{6};
  \arrj{a3}{a1}{1};
  \arrj{a4}{a1}{1};
  \arrj{a5}{a1}{2};
  \arrja{a5}{a4}{2};
  \arrja{a4}{a3}{4};
  \arrj{a5}{a7}{2};
  \arrjd{a7}{a1}{1};
  \arrj{a7}{a0}{3};
  \arrj{a6}{a0}{2};
  \arrj{a6}{a2}{2};
  \arrj{a4}{a7}{1};
  \arrjb{a6}{a7}{2};
  \arrj{a3}{a7}{1};
  \draw[every node/.style={draw}] (10,30) node {45};
\end{tikzpicture}
\end{center}
\caption{The five resolutions of
  $\C^3/\Z_{11}$.}\label{fig:Z11}
\end{figure}
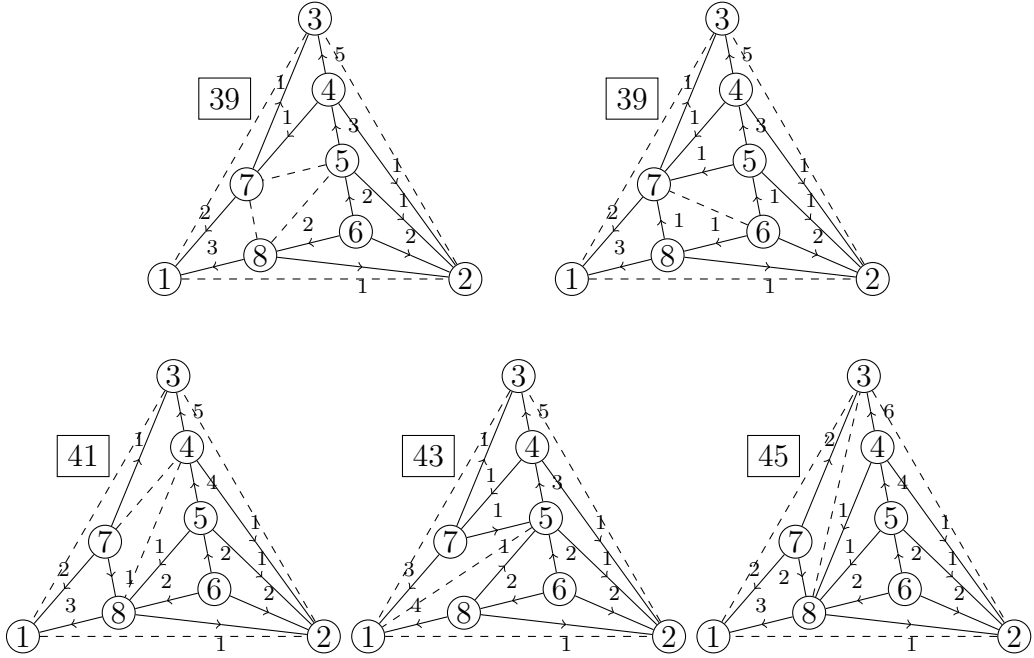

Now consider the geometric computation. There are 5 different possible
crepant resolutions of this orbifold all of which are
projective. These, and the associated values of $m_{ij}$ are shown in
figure \ref{fig:Z11}.\footnote{This figure also appeared in
  \cite{me:02mc} but the numbers have changed slightly due to a change
  in some definitions.}  The interior dotted lines represent
$m_{ij}=0$, i.e., $(-1,-1)$-curves. The border of the triangle is
shown as dotted as that also makes no contribution to the count. The
number appearing in a box to the left of each resolution is the total
count (\ref{eq:geomcount}).

So, the first two resolutions in figure \ref{fig:Z11} are in agreement
with conformal field theory while the other three have an excess in
the geometric count. This difference must be due to worldsheet
instanton effects.

Note that the first of the resolutions in figure \ref{fig:Z11} is the
distinguished resolution of the Hilbert scheme. This always has a
minimal number of (0,2)-deformations and does not require any
instanton corrections to agree with conformal field theory
\cite{BenG:Hilb}. We will prove that no toric irreducible rational
curves give instanton corrections to the Hilbert scheme in section
\ref{ss:nonisol}. So conformal field theory and geometry agree nicely
in this case.

\subsection{Deviations from the Hilbert Scheme}  \label{ss:dev}

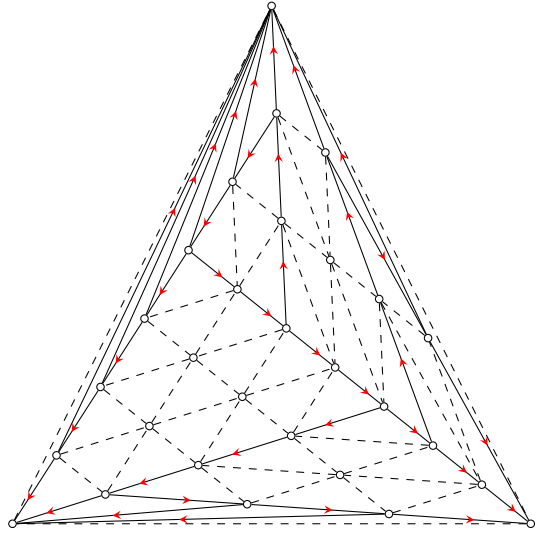
\begin{figure}
\begin{center}
\begin{tikzpicture}[x=1.3mm,y=1.3mm]
\path[shape=circle,inner sep=1pt,every node/.style={draw,font=\tiny}]
  (0,0) node(a0) {} 
  (26.5,53) node(a1) {} 
  (53,0) node(a2) {} 
  (4.5,7) node(a3) {} 
  (9.0,14) node(a4) {} 
  (13.5,21) node(a5) {} 
  (18.0,28) node(a6) {} 
  (22.5,35) node(a7) {} 
  (27.0,42) node(a8) {} 
  (9.5,3) node(a9) {} 
  (14.0,10) node(a10) {} 
  (18.5,17) node(a11) {} 
  (23.0,24) node(a12) {} 
  (27.5,31) node(a13) {} 
  (32.0,38) node(a14) {} 
  (19.0,6) node(a15) {} 
  (23.5,13) node(a16) {} 
  (28.0,20) node(a17) {} 
  (32.5,27) node(a18) {} 
  (24.0,2) node(a19) {} 
  (28.5,9) node(a20) {} 
  (33.0,16) node(a21) {} 
  (37.5,23) node(a22) {} 
  (33.5,5) node(a23) {} 
  (38.0,12) node(a24) {} 
  (42.5,19) node(a25) {} 
  (38.5,1) node(a26) {} 
  (43.0,8) node(a27) {} 
  (48.0,4) node(a28) {}; 
\draw[postaction={on each segment={mid arrow=red}}] (a1) -- (a25); 
\draw[postaction={on each segment={mid arrow=red}}] (a1) -- (a14) -- (a22) -- (a27);;
\draw[postaction={on each segment={mid arrow=red}}] (a1) -- (a8) -- (a13) -- (a17);
\draw[postaction={on each segment={mid arrow=red}}] (a1) -- (a7);
\draw[postaction={on each segment={mid arrow=red}}] (a1) -- (a6);
\draw[postaction={on each segment={mid arrow=red}}] (a1) -- (a5);
\draw[postaction={on each segment={mid arrow=red}}] (a1) -- (a4);
\draw[postaction={on each segment={mid arrow=red}}] (a1) -- (a3);
\draw[postaction={on each segment={mid arrow=red}}] (a0) -- (a3) -- (a4) -- (a5) -- (a6) -- (a7) -- (a8);  
\draw[postaction={on each segment={mid arrow=red}}] (a0) -- (a9) -- (a15) -- (a20) -- (a24);
\draw[postaction={on each segment={mid arrow=red}}] (a0) -- (a19);
\draw[postaction={on each segment={mid arrow=red}}] (a0) -- (a26);
\draw[postaction={on each segment={mid arrow=red}}] (a2) -- (a26) -- (a19) -- (a9); 
\draw[postaction={on each segment={mid arrow=red}}] (a2) -- (a28) -- (a27) -- (a24) -- (a21) -- (a17) -- (a12) -- (a6);
\draw[postaction={on each segment={mid arrow=red}}] (a2) -- (a25) -- (a14);

\draw[dashed] (a14) -- (a18) -- (a21);
\draw[dashed] (a22) -- (a24);
\draw[dashed] (a13) -- (a21);
\draw[dashed] (a8) -- (a14);
\draw[dashed] (a8) -- (a18) -- (a24);
\draw[dashed] (a13) -- (a18) -- (a22);
\draw[dashed] (a22) -- (a28) -- (a25) -- (a22);
\draw[dashed] (a7) -- (a12) -- (a13) -- (a7);
\draw[dashed] (a3) -- (a9) -- (a10) -- (a15) -- (a16) -- (a20) --
(a21);
\draw[dashed] (a3) -- (a10) -- (a16) -- (a21);
\draw[dashed] (a4) -- (a10) -- (a11) -- (a16) -- (a17) -- (a11) --
(a4);
\draw[dashed] (a5) -- (a11) -- (a12) -- (a5);
\draw[dashed] (a20) -- (a23) -- (a27) -- (a20);
\draw[dashed] (a15) -- (a19) -- (a23) -- (a26) -- (a28) --
  (a23) -- (a15);
\draw[dashed] (a0) -- (a1) -- (a2) -- (a0);
\end{tikzpicture}
\end{center} \caption{Hilbert Scheme Resolution for $G=\Z_{53}$.}
\label{fig:53}
\end{figure}

To get a better idea of the Hilbert scheme we show the case
$G\cong\Z_{53}$, with $(\nu_1,\nu_2,\nu_3)=
(\ff1{53},\ff7{53},\ff{45}{53})$, in figure
\ref{fig:53}. We do not show the values of $m$ to avoid cluttering the
diagram. The general following structure was proved in 
\cite{MR2075608}:\footnote{The arrows as defined in \cite{MR2075608}
  point in the opposite directions to ours. We choose our orientation
  to coincide with the directions of associated Ext groups.}
\begin{prop} \label{prop:Craw}
If we draw a triangulation $\Delta$ corresponding to the Hilbert
scheme of $\C^3/G$ using arrows and dotted lines as in section
\ref{ss:geom} then
\begin{enumerate}
\item There is a set of ``big'' triangles (of possibly non-minimal
  area) with non-dotted edges. These triangles are ``equilateral'' with
  respect to counting lattice points in the edges. The arrows on the
  edges of these triangles all point towards an outer vertex where the
  line terminates.
\item Each of the above big equilateral triangles is subdivided, if necessary,
  with dotted lines into a set of similar triangles of minimal area.
\end{enumerate}
\end{prop}

A nice systematic way to hunt for instanton effects is to deviate from
the Hilbert scheme. Consider $\C^3/\Z_n$ for a very large value of
$n$. The sizes of the ``big'' triangles depend on the choice of
weights but there will be a choice of weights for which at least one
of these triangles is very big. Then $\Delta$ for this Hilbert scheme
will contain a regular tessellation with a large number of small
triangles.  Within this regular tessellation the toric rational curves
are all of type $(-1,-1)$. Now perturb this regular tessellation by
performing a flop. Any
non-$(-1,-1)$-curves produced will add extra (0,2)-deformations and
thus must be associated to instanton effects.

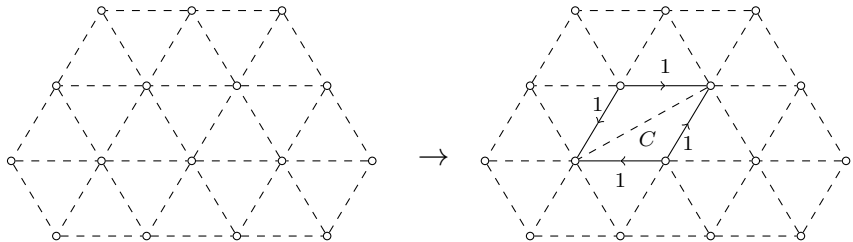
\begin{figure}[h]
\begin{center}
\begin{tikzpicture}[x=1.2mm,y=1.0mm]
\path[shape=circle,inner sep=1pt,
    every node/.style={draw,font=\tiny}]
  (0,20) node(a1) {}
  (10,20) node(a2) {}
  (20,20) node(a3) {}
  (-5,10) node(a4) {}
  (5,10)  node(a5) {}
  (15,10) node(a6) {}
  (25,10) node(a7) {}
  (-10,0) node(a8) {}
  (0,0) node(a9) {}
  (10,0) node(a10) {}
  (20,0) node(a11) {}
  (30,0) node(a12) {}
  (-5,-10) node(a13) {}
  (5,-10) node(a14) {}
  (15,-10) node(a15) {}
  (25,-10) node(a16) {};
\draw[dashed] (a3) -- (a2) -- (a1) -- (a5) -- (a10) -- (a6) -- (a5) --
(a9);
\draw[dashed] (a5) -- (a4) -- (a9) -- (a14) -- (a15) -- (a16) -- (a12)
  -- (a11) -- (a6) -- (a2) -- (a5);
\draw[dashed] (a4) -- (a8) -- (a9) -- (a13) -- (a14);
\draw[dashed] (a8) -- (a13);
\draw[dashed] (a4) -- (a1);
\draw[dashed] (a3) -- (a6) -- (a7) -- (a11);
\draw[dashed] (a3) -- (a7) -- (a12);
\draw[dashed] (a16) -- (a11) -- (a15) -- (a10) -- (a14);
\draw[dashed] (a9) -- (a10) -- (a11);
\end{tikzpicture}
\quad\raisebox{10mm}{$\to$}\quad
\begin{tikzpicture}[x=1.2mm,y=1.0mm]
\path[shape=circle,inner sep=1pt,
    every node/.style={draw,font=\tiny}]
  (0,20) node(a1) {}
  (10,20) node(a2) {}
  (20,20) node(a3) {}
  (-5,10) node(a4) {}
  (5,10)  node(a5) {}
  (15,10) node(a6) {}
  (25,10) node(a7) {}
  (-10,0) node(a8) {}
  (0,0) node(a9) {}
  (10,0) node(a10) {}
  (20,0) node(a11) {}
  (30,0) node(a12) {}
  (-5,-10) node(a13) {}
  (5,-10) node(a14) {}
  (15,-10) node(a15) {}
  (25,-10) node(a16) {};
\draw[dashed] (a3) -- (a2) -- (a1) -- (a5);
\draw[dashed] (a5) -- (a4) -- (a9) -- (a14) -- (a15) -- (a16) -- (a12)
  -- (a11) -- (a6) -- (a2) -- (a5);
\draw[dashed] (a4) -- (a8) -- (a9) -- (a13) -- (a14);
\draw[dashed] (a8) -- (a13);
\draw[dashed] (a4) -- (a1);
\draw[dashed] (a3) -- (a6) -- (a7) -- (a11);
\draw[dashed] (a3) -- (a7) -- (a12);
\draw[dashed] (a16) -- (a11) -- (a15) -- (a10) -- (a14);
\draw[dashed] (a10) -- (a11);
\draw[dashed] (a6) -- (a9);
\arrj{a5}{a9}{1};
\arrj{a5}{a6}{1};
\arrjd{a10}{a9}{1};
\arrjd{a10}{a6}{1};
\draw (8,3) node {$\scriptstyle C$};
\end{tikzpicture}
\end{center}
\caption{A simple lattice defect} \label{fig:defect1}
\end{figure}

This simplest such modification to form a lattice ``defect'' is shown in figure
\ref{fig:defect1}. Relative to the Hilbert scheme, it adds 4 to the
geometric count of deformations of $\cT$. Since this should have the
same number of (0,2)-deformations as the Hilbert scheme, all 4 of the
deformations of $\cT$ must be obstructed by instantons. We analyze
this in section \ref{sss:defect1}.


So if we have a single flop on the interior of a regular triangular lattice, it induces $4$ deformations 
that correspond to the instanton corrections.  What happens as we introduce more flops?
\begin{prop}
 If we take $n$ flops of a regular lattice structure, it adds between $4\sqrt{n}$ and $4n$ deformations of $\cT$ 
\end{prop}

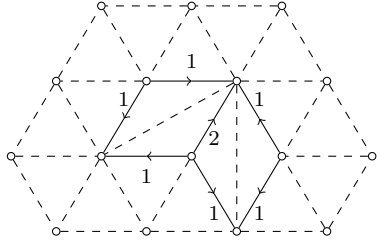
\begin{figure}
\begin{center}
\begin{tikzpicture}[x=1.2mm,y=1.0mm]
\path[shape=circle,inner sep=1pt,
    every node/.style={draw,font=\tiny}]
  (0,20) node(a1) {}
  (10,20) node(a2) {}
  (20,20) node(a3) {}
  (-5,10) node(a4) {}
  (5,10)  node(a5) {}
  (15,10) node(a6) {}
  (25,10) node(a7) {}
  (-10,0) node(a8) {}
  (0,0) node(a9) {}
  (10,0) node(a10) {}
  (20,0) node(a11) {}
  (30,0) node(a12) {}
  (-5,-10) node(a13) {}
  (5,-10) node(a14) {}
  (15,-10) node(a15) {}
  (25,-10) node(a16) {};
\draw[dashed] (a3) -- (a2) -- (a1) -- (a5);
\draw[dashed] (a5) -- (a4) -- (a9) -- (a14) -- (a15) -- (a16) -- (a12)
  -- (a11);
\draw[dashed] (a6) -- (a2) -- (a5);
\draw[dashed] (a4) -- (a8) -- (a9) -- (a13) -- (a14);
\draw[dashed] (a8) -- (a13);
\draw[dashed] (a4) -- (a1);
\draw[dashed] (a3) -- (a6) -- (a7) -- (a11);
\draw[dashed] (a3) -- (a7) -- (a12);
\draw[dashed] (a16) -- (a11);
\draw[dashed] (a10) -- (a14);
\draw[dashed] (a6) -- (a15);
\draw[dashed] (a6) -- (a9);
\arrj{a5}{a9}{1};
\arrj{a5}{a6}{1};
\arrjd{a10}{a9}{1};
\arrjd{a10}{a15}{1};
\arrjd{a10}{a6}{2};
\arrjd{a11}{a15}{1};
\arrj{a11}{a6}{1};

\end{tikzpicture}
\end{center}
\caption{Complimentary Lattice Defects} \label{fig:defect2}
\end{figure}

We first show that the most deformations that correspond to $n$ flops is $4n$.  Anytime we take a flop, it adds at most $4$ deformations (one for 
each curve that shares a triangle with the changed curve).  Thus, if each of $n$ flops adds $4$ deformations, we have $4n$ deformations.  There are 
two ways this can occur.   If all of the flops are isolated, we end up with $n$ parallelograms that look like figure \ref{fig:defect1}, where each 
arrow has multiplicity 1.  It is also possible that any number of the defects be adjacent, but changed in a complimentary way.  
An example where this occurs can be seen in figure \ref{fig:defect2}.  In this case every arrow has multiplicity 1 or 2, and the total number of deformations 
is still $4n$.

We now show that the lower bound on the number of deformations is given by $4\sqrt{n}$.  This bound is achieved when $\sqrt{n}$ is 
an integer, and all of the flops occur within a $\sqrt{n}$ by $\sqrt{n}$ quadrilateral.
Each time we take a flop, we want to minimize the number of deformations added.  This means we must have as many $(-1,-1)$-curves as possible.  
This is done by taking flops that border the deformations we already have (which correspond to $(-2,0)$-curves), so that they become $(-1,-1)$-curves 
again. If one such curve is changed in this way, then this flop adds two deformations (instead of the 4 we saw earlier).  
If two are changed, then this flop does not change the number of deformations.  
Each time a flop is added, use the one that changes as many curves in this way as is possible.  If we follow this algorithm, 
beginning with a single defect as in figure \ref{fig:defect1}, we spiral outward.  It is clear that this minimizes the total number of deformations.  We also see 
that whenever these flops occur in a $\sqrt{n}\times\sqrt{n}$ quadrilateral, there are a total of $4\sqrt{n}$ deformations.  Since this process 
adds 2 deformations with the next flop, and another 2 deformations halfway to the next perfect quadrilateral, the total number of deformations 
is always at least $4\sqrt{n}$.$\square$

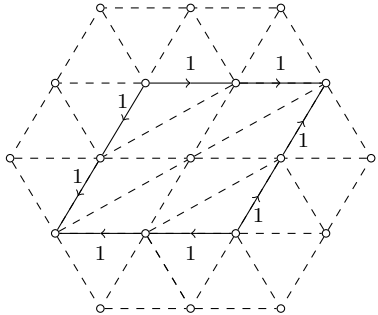
\begin{figure}
\begin{center}
\begin{tikzpicture}[x=1.2mm,y=1.0mm]
\path[shape=circle,inner sep=1pt,
    every node/.style={draw,font=\tiny}]
  (0,20) node(a1) {}
  (10,20) node(a2) {}
  (20,20) node(a3) {}
  (-5,10) node(a4) {}
  (5,10)  node(a5) {}
  (15,10) node(a6) {}
  (25,10) node(a7) {}
  (-10,0) node(a8) {}
  (0,0) node(a9) {}
  (10,0) node(a10) {}
  (20,0) node(a11) {}
  (30,0) node(a12) {}
  (-5,-10) node(a13) {}
  (5,-10) node(a14) {}
  (15,-10) node(a15) {}
  (25,-10) node(a16) {}
  (0,-20) node(a17) {}
  (10,-20) node(a18) {}
  (20,-20) node(a19) {};
\draw[dashed] (a3) -- (a2) -- (a1) -- (a5);
\draw[dashed] (a5) -- (a4) -- (a9);
\draw[dashed] (a14) -- (a15) -- (a16) -- (a12)
  -- (a11);
\draw[dashed] (a6) -- (a2) -- (a5);
\draw[dashed] (a4) -- (a8) -- (a9) -- (a13);
\draw[dashed] (a8) -- (a13) --(a17);
\draw[dashed] (a4) -- (a1);
\draw[dashed] (a14) -- (a13);
\draw[dashed] (a3) -- (a6) -- (a7) -- (a11)--(a15);
\draw[dashed] (a3) -- (a7) -- (a12);
\draw[dashed] (a16) -- (a11);
\draw[dashed] (a14) -- (a18);
\draw[dashed] (a15) -- (a19);
\draw[dashed] (a16) -- (a19);
\draw[dashed] (a15) -- (a18);
\draw[dashed] (a14) -- (a17);
\draw[dashed] (a17) -- (a18) -- (a19);
\draw[dashed] (a6) -- (a10) -- (a14);
\draw[dashed] (a6) -- (a9);
\draw[dashed] (a9) -- (a10) -- (a11);
\draw[dashed] (a13) -- (a10) -- (a7);
\draw[dashed] (a14) -- (a11);
\draw[dashed] (a14) -- (a18);
\arrj{a5}{a9}{1};
\arrj{a9}{a13}{1};
\arrjd{a14}{a13}{1};
\arrjd{a15}{a14}{1};
\arrj{a5}{a6}{1};
\arrj{a6}{a7}{1};
\arrjd{a11}{a7}{1};
\arrjd{a15}{a11}{1};

\end{tikzpicture}
\end{center}
\caption{Minimizing Lattice Defects} \label{fig:defect3}
\end{figure}

Figure \ref{fig:defect3} shows an example which achieves this lower bound, in the case where there are 4 total flops.  We note that there are no ``interior''
 deformations.  This is always the case when following the above minimization procedure.  We also note that all arrows have multiplicity 1.

Lattice defects within the regular tessellation always lead to an even number of deformations, and every even integer between $4\sqrt{n}$ and $4n$ 
can be achieved as the number of deformations by using a combination of the above methods.  
It is also possible to obtain arrows with multiplicities higher than two, through a combination of the methods described above.


\section{Worldsheet Instantons}  \label{s:inst}

For the supersymmetric non-linear $\sigma$-model with target space
$X$, the instantons are given by holomorphic maps from the string
worldsheet to $X$ \cite{Wen:1985jz}. In an instanton calculation the
path integral is reduced to an integral over the moduli space of the
instanton solutions. In the presence of fermions this will usually be
a supermoduli space. Thus, when computing an instanton correction to a
correlation functions, we get a zero value if the number of
fermions in the integrand is not equal to the number of fermions in the
supermoduli space measure. 

An alternative approach to computing instantons is to view one
intrinsically as a subspace of $X$ and use the physics of D-branes
\cite{W:K3inst}. When the contribution is nonzero, this gives a
slightly more practical methods for computing corrections to the
superpotential as done in \cite{Buchbinder:2002ic}, for example. The
disadvantage is that this requires the assumption that the instanton
image is a smooth curve. We will therefore use the older method
coming from conformal field theory.

Let $\phi$ be the $\sigma$-model map from the worldsheet to $X$.
We then have left-moving fermions taking values in
$K^{\frac12}\otimes\phi^*(V\oplus \overline{V})$, where $K$ is the canonical
bundle of the worldsheet, i.e., $K\cong\O(-2)$. Left-moving fermion
zero modes are holomorphic sections of this bundle:
\begin{equation}
\begin{split}
H^0(K^{\frac12}\otimes\phi^*(V)) &= H^0(\phi^*(V)\otimes\O(-1))\\
H^0(K^{\frac12}\otimes\phi^*(\overline{V})) &= H^1(\phi^*(V)\otimes\O(-1)),
\end{split} \label{eq:lambda0}
\end{equation}
where the second line follows from Serre duality and $\bar V\cong
V^\vee$. The right-moving fermions take values in $\bar
K^{\frac12}\otimes(\cT\oplus\overline{\cT})$. The zero modes are
antiholomorphic sections of this bundle. The easiest thing to do is to
take the complex conjugate and look at holomorphic sections
again. Thus the zero modes are
\begin{equation}
\begin{split}
H^0(K^{\frac12}\otimes\phi^*(\cT)) &= H^0(\phi^*(\cT)\otimes\O(-1))\\
H^0(K^{\frac12}\otimes\phi^*(\overline{\cT})) &= H^1(\phi^*(\cT)\otimes\O(-1)).
\end{split}
\end{equation}

For this section we suppose the instanton is an {\em embedding} of the
worldsheet with image a rational curve $C\subset X$. Then we may
identify $\phi^*(V)$ with the restriction $V|_C$. All vector
bundles over a rational curve split into a sum of line bundles and $V$
is rank three.
Thus
\begin{equation}
  \left.V\right|_C = \O(a)\oplus\O(b)\oplus\O(c),  \label{eq:split2}
\end{equation}
where $a+b+c=c_1(V)=0$.  The number of left-moving fermion zero modes,
from (\ref{eq:lambda0}), is $|a|+|b|+|c|$.

Now let us consider the right-moving fermions zero modes by replacing
$V$ with $\cT$. The adjunction formula tells us there is an exact sequence
\begin{equation}
\xymatrix@1{0\ar[r]&\cT_C\ar[r]&\cT|_C\ar[r]&\cN\ar[r]&0,} \label{eq:adj}
\end{equation}
where $\cT_C\cong\O(2)$. This exact sequence need not split. Indeed we
will see an example in section \ref{ss:nontoric} where $\cT|_C$ does
not include a $\O(2)$ factor. Having said that, we will show in
section \ref{ss:rst} that (\ref{eq:adj}) always splits when $C$ is
toric.

Since $\cN\cong\O(k)\oplus\O(l)$ we see $C$ is a
$(k,l)$-curve. Clearly $k+l=-2$. If neither $k$ nor $l$ is greater than
3 then $\Ext^1(\cN,\cT_C)=0$ and so (\ref{eq:adj}) splits again.  A
first-order deformation of $C\subset X$ corresponds to a section of
$\cN$. Thus, the case $k=l=-1$ has no deformations and the
$(-1,-1)$-curve $C$ is isolated. The converse is not, in general,
true. Any curve other than $(-1,-1)$ will have first order
deformations but these deformations may be obstructed and thus the
curve may still be isolated. See \cite{me:13quiv} for a detailed
example. Luckily, toric varieties do not suffer from such
obstructions. That is because the normal bundle of $C$ is explicitly
a Zariski open set in $X$. Thus, in this paper, rational curves are isolated
if and only if they are $(-1,-1)$-curves.

The case considered in \cite{DSWW:} was for a $(-1,-1)$-curve. The
right-moving sector, where the fermions take values in
$\cT|_C=\O(2)\oplus\O(-1)\oplus(-1)$ therefore has 4 zero modes. If
$V$ is suitably generic then $a=b=c=0$ and thus there are no
left-moving zero modes. It is precisely this case where the fermionic
dimensions work out to give a potentially nonzero instanton correction
to the superpotential. These theories are therefore ``bad''. A first
order deformation of the tangent bundle to such a $V$ should not
correspondence to a massless state.

In this paper we are concerned only with the question of whether instanton
corrections are nonzero or not. To actually compute the magnitude it
is necessary to compute Pfaffians and determinants. See
\cite{Buchbinder:2002ic} for an example. Here we note that, at least
in the case of a smooth $(-1,-1)$-curve, the contribution is nonzero
if and only if the fermion zero mode count is as above.

If $C$ is not a $(-1,-1)$-curve it will move in a family. This may, or
may not lead to extra right-moving fermion zero modes. If it does,
this will change the condition for instanton corrections for $V$ to
something other than the trivial splitting $a=b=c=0$. We will discuss
this in section \ref{ss:nonisol}.


\section{Splitting}   \label{s:split}

\subsection{Deformations of $\cT$.}  \label{ss:defs}
Let $\mathbf{q}_i$ denote the multi-degree, or ``charge'', of
$x_i$. It was shown in \cite{BatCox:toric} that the tangent sheaf
$\cT$ of $X_\Sigma$ is given by
\begin{equation}
\xymatrix@1@C=15mm{
0\ar[r]&\O^{\oplus r}\ar[r]^-E&\bigoplus_{i=1}^n \O(\mathbf{q}_i)
\ar[r]&\cT\ar[r]&0,
} \label{eq:torictan}
\end{equation}
where $E$ is an $n\times r$ matrix whose $(i,j)$-th entry is
$\Phi_{ji}x_i$ and $\Phi$ is the matrix in (\ref{eq:toric}).
It is important to note that each summand in middle term of
(\ref{eq:torictan}) corresponds to a point in $\cA$ and thus a point in
our toric diagrams. This is central to way we will visualize the
constructions below.

It follows from (\ref{eq:torictan}) that $\sHom(\cT,\cT)$ is
isomorphic in $\DC(X)$ to the complex
\begin{equation}
\xymatrix@1{0\ar[r]&\bigoplus_i \O(-\mathbf{q}_i)^{\oplus
    r}\ar[r]&{\begin{matrix}\O^{\oplus r^2}\\
  \oplus\\
  \bigoplus_{i,j}\O(\mathbf{q}_i-\mathbf{q}_j)\end{matrix}}\ar[r]&
  \bigoplus_i \O(\mathbf{q}_i)^{\oplus
    r}\ar[r]&0.
}
\end{equation}
We can then use the hypercohomology spectral sequence to compute the
space of first order deformations of $\cT$ which is $\Ext^1(\cT,\cT)=
H^1(X,\sHom(\cT,\cT))$. We are doing a simpler version of the
computation that appeared in \cite{meMP:singlets,AP:elusive}.

Note that many of the cohomology groups vanish:
\begin{prop}  \label{prop:vanish}
Let $\alpha$ be a vertex of $\Delta$ and let $D_\alpha$ be
the associated toric divisor. Then
\begin{equation}
\begin{split}
  H^i(\O_{D_\alpha}(\mathbf{q_\alpha})) &= 0, \qquad \hbox{for
    $i=0,1$, $\alpha$ interior}\\
  H^1(\O(\mathbf{q_\alpha})) &= 0,\\
  H^i(\O(-\mathbf{q_\alpha})) &= 0, \qquad \hbox{for $i=1,2.$}\\
\end{split}
\end{equation}
\end{prop}

To prove the first line note that $\O_{D_\alpha}(-\mathbf{q_\alpha})$
is the conormal bundle of $D_\alpha$. Thus, since $X$ is \CY, by
adjunction we have that $\O_{D_\alpha}(\mathbf{q_\alpha})$ is the
canonical bundle of $D_\alpha$. The fact that $\alpha$ is an interior
point means that $D_\alpha$ is compact. Now use Serre duality and the fact
that $H^i(\O_{D_\alpha})=0$ for $i>0$, since $D_\alpha$ is a rational surface.
If $\alpha$ is interior then other two
results follow from the first result and the short exact sequences
\begin{equation}
\xymatrix@1{0\ar[r]&\O(\mathbf{a})\ar[r]&
  \O(\mathbf{a+\mathbf{q_\alpha}})\ar[r]&
  \O_{D_\alpha}(\mathbf{q_\alpha})\ar[r]&0
},
\end{equation}
for $\mathbf{a}=0$ and $-\mathbf{q_\alpha}$ respectively. If $\alpha$
is a vertex of the triangle then $\O(\mathbf{q_\alpha})$ is a
``tautological line bundle'' in the McKay correspondence as in
\cite{GS-V:Mckay}. These, and $\O$, are projective objects in the
category of coherent sheaves and thus the required cohomology groups
vanish.
\QED

The spectral sequence is therefore quite easy to compute and the
result is that we have two contributions to $\Ext^1(\cT,\cT)$ with
obvious interpretations:
\begin{enumerate}
\item $\bigoplus_iH^0(\bigoplus_i \O(\mathbf{q}_i)^{\oplus r})$ contributes
  deformations of the map $E$ in (\ref{eq:torictan}).\footnote{Some of
  these are killed by the $d_1$ differential in the spectral
  sequence. This amounts to subtracting simple reparametrizations from
  the count as explained in \cite{meMP:singlets}. Care must be taken
  in this step to count {\em framed\/} deformations. This is done
  in \cite{me:02mc}.}
\item $\bigoplus_{i,j}H^1(\O(\mathbf{q}_i-\mathbf{q}_j))$ gives
  deformations of the direct sum in the middle of (\ref{eq:torictan}).
\end{enumerate}

The first type of deformations, the deformations of $E$, should be
viewed as the ``easy'' deformations. They have been much-studied in
the context of the gauged linear $\sigma$-model
\cite{Beasley:2003fx,McOrist:2008ji,Kreuzer:2010ph}. These do not
depend on the choice of resolution and we will see in section
\ref{ss:rest} that they never suffer from instanton corrections. The
second type of deformations are the focus of this paper.

\subsection{Restricting the Tangent Bundle} \label{ss:rst}

Let us restrict the presentation of $\cT$ from (\ref{eq:torictan}) to
the $(-1-m,-1+m)$ curve $C$ appearing in figure \ref{fig:C}. This
curve is given by $x_i=x_j=0$ with homogeneous coordinate
$[x_\alpha,x_\beta]$. Let $\zeta_k$ denote one of the $n-4=r-1$ elements
of $\cA$ not appearing in figure \ref{fig:C} and let $x_{\zeta_k}$ be
the associated coordinate. There is only one $\C^*$-action in
$(\C^*)^r$ that leaves all the $x_{\zeta_k}$'s invariant. This comes
from the affine relation between the 4 points in figure \ref{fig:C}
and yields a charge $Q$ given by:
\begin{equation}
\begin{array}{c|cccc}
&x_\alpha&x_\beta&x_i&x_j\\
\hline
Q&1&1&-1-m&-1+m
\end{array}
\end{equation}
It follows that tangent sheaf restricts to $C$ as the cokernel of the map
\begin{equation}
\xymatrix@1{\O^{\oplus r}\ar[r]^-E&
\O(1)^{\oplus 2}\oplus\O(-1-m)\oplus\O(-1+m)
\oplus\O^{\oplus r-1},}  \label{eq:Tan2}
\end{equation}
where $\O$ is now, of course, the structure sheaf of $C$.
Now, since $\{i,j,\zeta_k\}$ do not form the vertices of a simplex in
the triangulation of $\Delta$, $(x_i,x_j,x_{\zeta_k})\supset B$. Thus,
on $C$ where $x_i=x_j=0$ we must have $x_{\zeta_k}\neq0$. This gives a
great deal of redundancy in the presentation (\ref{eq:Tan2}). We may
cancel $r-1$ of the $\O$'s to give
\begin{equation}
\xymatrix@1@C=15mm{
0\ar[r]&\O\ar[r]^-{\left(\begin{smallmatrix}x_\alpha\\x_\beta\\0\\0
\end{smallmatrix}\right)}&
{\begin{matrix}\O(1)^{\oplus 2}\\\oplus\\\O(-1-m)\\\oplus\\\O(-1+m)
\end{matrix}}\ar[r]&\cT|_C\ar[r]&0.} \label{eq:Tan10}
\end{equation}
But there is a short exact sequence on $C$:
\begin{equation}
\xymatrix@1@C=15mm{0\ar[r]&\O\ar[r]^-{\left(\begin{smallmatrix}x_\alpha\\x_\beta
\end{smallmatrix}\right)}&\O(1)^{\oplus 2}
\ar[r]^-{\left(\begin{smallmatrix}-x_\beta&x_\alpha\end{smallmatrix}\right)}
&\O(2)\ar[r]&0,}
\end{equation}
from which it follows that
\begin{equation}
  \cT|_C = \O(2)\oplus\O(-1-m)\oplus\O(-1+m),
\end{equation}
as promised.

\subsection{Restricting the Deformation}  \label{ss:rest}

In this section we will restrict a deformation of the tangent bundle
to a toric rational curve $C\subset X$. First we prove the following,
which will be enough to prove that worldsheet instantons never
obstruct the ``easy'' deformations of $\cT$.
\begin{prop}
  First order deformations of $\cT$ coming from first order
  deformations of the map $E$ in (\ref{eq:torictan}) will always
  give $\O(-1-m)$ as a summand when restricted to $C$.
\end{prop}
To see this note that the fact we consider only small deformations
means that we assume the nonzero entries in $E$ coming from the
condition $x_{\zeta_k}\neq0$ remain nonzero and we may still cancel
the $r-1$ $\O$'s to get (\ref{eq:Tan10}). Furthermore, since
$x_\alpha$ and $x_\beta$ have positive degree, we cannot have any
nonzero expression in the third entry of the matrix appearing over the first
map in (\ref{eq:Tan10}). Therefore, the $\O(-1-m)$ summand must be
preserved unchanged in the cokernel of this map. \QED

So now we turn to the ``hard'' deformations associated to
$H^1(\O(\mathbf{q_\beta}-\mathbf{q_\gamma}))$.  We can study these
very explicitly using the local cohomology picture of sheaf cohomology
in toric varieties from \cite{EMS:ToricCoh}.

\begin{figure}
\begin{center}
\begin{tikzpicture}[x=0.3mm,y=0.3mm]
  \path[shape=circle,inner sep=1pt,minimum size=5mm,every node/.style={draw}]
     (0,0) node(a0) {$\alpha$}
     (50,50) node(a1) {$\beta$}
     (0,50) node(a2) {$\gamma$}
     (50,0) node(a3) {$\delta$}
     (-30,100) node(a4) {$\epsilon$};
  \draw[dashed] (a0) -- (a1);
  \arrj{a2}{a1}{m};
  \draw (a0) -- (a2) -- (a4) -- (a1) -- (a3) -- (a0);
  \draw(15,44) node {$\scriptstyle T$};
  \draw(25,18) node {$\scriptstyle C$};
\end{tikzpicture}
\end{center}
\caption{Part of $\Delta$ used in restricting deformations.} \label{fig:part}
\end{figure}
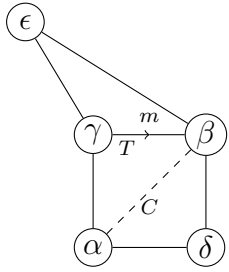

For the sake of exposition we will assume that $C$ is a
$(-1,-1)$-curve. Then let us restrict attention to an internal part of
the triangulation of $\Delta$ that looks like figure~\ref{fig:part}.
Let $x_\alpha,\ldots,x_\epsilon$ be the corresponding homogeneous
variables. $C$ is then the $(-1,-1)$-curve given by
$x_\alpha=x_\beta=0$ and let $T$ be the $(-1-m,-1+m)$-curve given by
$x_\beta=x_\gamma=0$. We are going to be interested in worldsheet
instantons over $C$ associated to hard
$H^1(\O(\mathbf{q_\beta}-\mathbf{q_\gamma}))$ deformations originating
from $T$. We will first confirm the result stated earlier that this
contributes $m$ first order deformations of the tangent bundle.

From the sequence,
\begin{equation}
\xymatrix@1{0\ar[r]&\O(-\mathbf{q_\gamma})
  \ar[r]^-{x_\beta}&\O(\mathbf{q_\beta}-\mathbf{q_\gamma})\ar[r]&
  \O_{D_\beta}(\mathbf{q_\beta}-\mathbf{q_\gamma})\ar[r]&0,
}
\end{equation}
 it follows from proposition \ref{prop:vanish} that 
\begin{equation}
H^1(\O(\mathbf{q_\beta}-\mathbf{q_\gamma}))\cong
H^1(\O_{D_\beta}(\mathbf{q_\beta}-\mathbf{q_\gamma})).
\end{equation}
 Furthermore,
using
\begin{equation}
\xymatrix@1{0\ar[r]&\O_{D_\beta}(\mathbf{q_\beta}-\mathbf{q_\gamma})
  \ar[r]^-{x_\gamma}&\O_{D_\beta}(\mathbf{q_\beta})\ar[r]&
  \O_T(\mathbf{q_\beta})\ar[r]&0}  \label{eq:DtoT}
\end{equation}
and proposition \ref{prop:vanish} we see that
$H^1(\O_{D_\beta}(\mathbf{q_\beta}-\mathbf{q_\gamma}))\cong
H^0(\O_T(\mathbf{q_\beta}))$. Thus we may explicitly count the desired
deformations by analyzing holomorphic functions on $T$ with the
correct charge. As in section \ref{ss:rst}, let $\zeta_k$ denote a
vertex in $\Delta$ that does not appear in
figure~\ref{fig:part}. So $x_{\zeta_k}\neq0$ on $T$ and
we may use the $(\C^*)^r$-action to fix $x_{\zeta_k}=1$. That is, we only
consider functions of $x_\alpha,\ldots,x_\epsilon$. Two of the
$(\C^*)^r$ actions have charges
\begin{equation}
\begin{array}{c|ccccc}
&x_\alpha&x_\beta&x_\gamma&x_\delta&x_\epsilon\\
\hline
Q_1&-1&-1&1&1&0\\
Q_2&1&-1-m&-1+m&0&1
\end{array}
\end{equation}
with charge zero for every vertex not appearing in
figure~\ref{fig:part}.

These charges can be used to determine a basis of monomials for
$H^0(\O_T(\mathbf{q_\beta}))$, i.e., holomorphic functions on $T$ with
the same charge as $x_\beta$. Obviously we can't use $x_\beta$ and
$x_\gamma$ since they are zero on $T$. The coordinates $x_\alpha$ and
$x_\epsilon$ have zeros on $T$ but $x_\delta$ does not. Thus, we can
have non-negative powers of $x_\alpha$ and $x_\epsilon$, and any power
of $x_\delta$. A basis is thus given by the $m$ monomials
\begin{equation}
x_\alpha^{m-1}x_\delta^{m-2},\; x_\alpha^{m-2}x_\epsilon x_\delta^{m-3},\;
x_\alpha^{m-3}x_\epsilon^2 x_\delta^{m-4},\ldots,
x_\alpha x_\epsilon^{m-2},\; x_\epsilon^{m-1}x_\delta^{-1}. \label{eq:mlist}
\end{equation}
So, as promised, we contribute $m$ first order deformations of the
tangent bundle. Furthermore, we have an explicit realization of these
deformations which allows us to restrict to $C$.

First map the holomorphic functions in $H^0(\O_T(\mathbf{q_\beta}))$
to rational functions representing local cohomology associated to
$H^1(\O_{D_\beta}(\mathbf{q_\beta}-\mathbf{q_\gamma}))$ using the
connecting homomorphism in the long exact sequence associated to 
(\ref{eq:DtoT}). This is clearly given by division by $x_\gamma$. Then
we restrict to $C$ by setting $x_\alpha=x_\beta=0$. Finally we may set
$x_\epsilon=1$ by using a $\C^*$-action since $x_\epsilon\neq0$ on
$C$. Thus the restriction map sends all the monomials in
(\ref{eq:mlist}) to zero except the final one which becomes
\begin{equation}
  \frac1{x_\gamma x_\delta}.
\end{equation}
This is precisely the monomial which in local cohomology represents
$H^1(C,\O(-2))\cong\C$. Therefore, restriction gives us a surjective map
\begin{equation}
\rho_C:H^1(\O_{D_\beta}(\mathbf{q_\beta}-\mathbf{q_\gamma})) \to
H^1(C,\O(-2)).
\end{equation}

This deformation comes from
$\Ext_X^1(\O(\mathbf{q_\gamma}),\O(\mathbf{q_\beta}))$ which restricts to
$\Ext_C^1(\O(1),\O(-1))$. The nontrivial extension of $\O(1)$ by
$\O(-1)$ on $\P^1$ is $\O\oplus\O$ (see appendix). This is how we
deform $\cT|_C$ in (\ref{eq:Tan2}).

\subsection{Examples}

\subsubsection{The simple defect case} \label{sss:defect1}

Let us apply the above result to the simple lattice defect in figure
\ref{fig:defect1}. Here we have a $(-1,-1)$-curve labeled $C$ which
intersects four $(-2,0)$-curves. The worldsheet instanton will be
associated to $C$ and the extra deformations of $\cT$, which we need
to kill, are associated to these four $(-2,0)$-curves. Let $[x,y]$ be
the homogeneous coordinates on $C$.

This is the case $m=1$ of the previous section. Each arrow in figure
\ref{fig:defect1} is associated to $\Ext_C^1(\O(1),\O(-1))\cong\C$. In terms
of $\cT|_C$, these can be viewed as the four dotted lines in:
\begin{equation}
\xymatrix@C=15mm@R=1mm{
&&\O(1)\ar@{.>}@/^8mm/@<5mm>[dddddd]\ar@{.>}@/_6mm/@<-5mm>[dddd]\\
&&\oplus\\
&&\O(1)\ar@{.>}@/^3mm/@<3mm>[dd]\ar@{.>}@/_6mm/@<-5mm>[dddd]\\
0\ar[r]&\O\ar[r]^-{\left(\begin{smallmatrix}x\\y\\0\\0
\end{smallmatrix}\right)}&
\qquad\quad\oplus\qquad\quad\ar[r]&\cT|_C\ar[r]&0,\\
&&\O(-1)\\
&&\oplus\\
&&\O(-1)\\
} \label{eq:Tan11}
\end{equation}
where each dotted line wants to turn $\O(1)\oplus\O(-1)$ into $\O\oplus\O$.

To analyze this more carefully we will rewrite this presentation of
$\cT|_C$ such that these deformations can be written simply as
deformations of maps in the complex. Replace each $\O(1)$ by the
quasi-isomorphic $\O(-1)\to\O^{\oplus2}$ (as in the appendix) and
reduce the presentation to get
\begin{equation}
\xymatrix@1@C=12mm{0\ar[r]&
\O(-1)^{\oplus2}\ar[r]^-{\left(\begin{smallmatrix}x&0\\y&x\\0&y\\
a&b\\c&d\end{smallmatrix}
\right)}&{\begin{matrix}\O^{\oplus3}\\\oplus\\\O(-1)^{\oplus2}\end{matrix}
}\ar[r]&\cT|_C\ar[r]&0,}
\end{equation}
where $a=b=c=d=0$. Now we see the 4 deformations explicitly by turning
on $a,b,c,d$. 

It immediately follows that $\cT$ deforms to $\cE$, where
\begin{equation}
\cE|_C = \begin{cases}
\O(2)\oplus\O(-1)\oplus\O(-1)\qquad&\hbox{if}\quad a=b=c=d=0,\\
\O(1)\oplus\O\oplus\O(-1)\qquad&\hbox{if otherwise and}\quad ad-bc=0,\\
\O\oplus\O\oplus\O\qquad&\hbox{otherwise}.
\end{cases}
\end{equation}

There is therefore good news and bad news. The good news is that for a
{\em generic\/} value of $a,b,c,d$ we do indeed get a trivially split bundle
and so instanton corrections prevent this being a good conformal field
theory. The bad news is that we are left with a 3-dimensional space 
$ad-bc=0$ where the bundle splits nontrivially. So we only removed one
of our 4 spurious deformations.

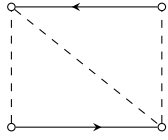
\begin{figure}
\begin{center}
\begin{tikzpicture}[x=1.0mm,y=0.8mm]
\path[shape=circle,inner sep=1pt,
    every node/.style={draw,font=\tiny}]
  (0,0) node(a1) {}
  (20,0) node(a2) {}
  (0,20) node(a3) {}
  (20,20) node(a4) {};
\draw[dashed] (a1) -- (a3) -- (a2) -- (a4);
\draw[postaction={on each segment={mid arrow}}] (a2) -- (a1);
\draw[postaction={on each segment={mid arrow}}] (a3) -- (a4);
\end{tikzpicture}
\end{center}
\caption{Minimal arrows for a $(-1,-1)$-curve instanton.} \label{fig:min11}
\end{figure}

\subsubsection{Other Hilbert Scheme Deviations}

The analysis of the previous section applies to any
$(-1,-1)$-curve. We get a very similar result in the case of
complimentary defects as in figure \ref{fig:defect2}. For every 4
obstructions we would like to obtain, we only get one.

In order to switch on Ext's to make the bundle
trivially split we need to make $ad-bc\neq0$. Thus we need, at a
minimum, arrows of the form of figure \ref{fig:min11}.
This provides a serious problem for the larger ``minimal'' defects as in
figure \ref{fig:defect3}. We are unable to get any instanton
obstructions whatsoever from irreducible $(-1,-1)$-curves.

\subsubsection{The $\Z_{11}$ case} \label{sss:Z11}

Looking at figure \ref{fig:Z11} we see a similar picture. We take each
of the five pictures in turn from top left to bottom right:
\begin{enumerate}
\item This is the Hilbert scheme.
\item At first sight the single $(-1,-1)$-curve looks similar to the
  case of the previous section. However, the arrows are not oriented
  correctly. In particular node 6 has arrows only pointing away from
  it. This means the $\O(-1)$ associated to node 6 can never be turned
  into $\O$. Thus we get no instanton corrections, as expected.
\item The $(-1,-1)$ curve joining nodes 4 and 8 gives instantons. One
  of the 4 arrows is missing compared to section \ref{sss:defect1} so
  we set $d=0$. So we have no instanton corrections if $bc=0$. This is
  a codimension one condition, i.e., this instanton only obstructs one
  deformation. We were hoping to obstruct $41-39=2$.
\item The $(-1,-1)$-curve joining nodes 1 and 5 is very similar to
  section \ref{sss:defect1}. We expect to kill 4 deformations but the
  $ad-bc=0$ condition results in only one deformation being lost.
\item The $(-1,-1)$-curve joining nodes 3 and 8 is again very similar to
  section \ref{sss:defect1}. So now we expect to kill 6 deformations
  but only end up killing one.
\end{enumerate}

The result is that in the last three cases we do not kill enough
deformations for geometry to agree with conformal field theory.


\section{Other Sources of Instantons} \label{s:oth}

\subsection{Non-isolated toric curves} \label{ss:nonisol}

We now turn our attention to rational curves other than
$(-1,-1)$-curves. In general there will be more right-moving zero
modes and left-moving zero modes. That is, we have a fermionic part of
the instanton moduli space over which we must integrate.

This fermionic integration is familiar from topological field theory
\cite{W:matrix,AM:rat} associated to the $N=(2,2)$ case. In this
$\sigma$-model, there is a term in the action given by
\begin{equation}
   R_{i\bar\imath j\bar\jmath} \lambda^i\bar\lambda^{\bar i}\psi^j\bar\psi^{\bar j},
  \label{eq:R4}
\end{equation}
where $\lambda$ and $\psi$ are the left and right-moving fermions
respectively, and $R_{i\bar\imath j\bar\jmath}$ is the Riemann
curvature of the \CY\ $X$. The integrand over the supermoduli space
contains the exponential of the action and thus any power of
(\ref{eq:R4}) appears. This allows an instanton contribution from any
extra even number of $\psi$ zero-modes so long as there is an equal
number of extra $\lambda$ zero-modes. In the topological field theory
case, something very beautiful happens. These powers of the Riemann
curvature tensor conspire to form the top Chern class which can be
integrated over the the moduli space to form the Euler characteristic
of the moduli space. 

In the (0,2) case we replace the Riemann curvature tensor with the
curvature of the bundle $V$, but in this paper we can put $V\cong\cT$
to first order anyway. However, we cannot expect things to be as nice
as the topological field theory case and should not expect to get
characteristic classes. In this paper we will not attempt to do the
instanton moduli space integration. We will just count fermion zero
modes to account for the {\em possibility\/} of instanton corrections.

Let $C$ have normal bundle $\O(-1-m)\oplus\O(-1+m)$. There are thus
$2+|m+1|+|m-1|$ right-moving fermion zero modes. To obtain a nonzero
instanton correction to the destabilizing one-point function we need
four more right-moving modes than left-moving modes. Thus we need
$|m+1|+|m-1|-2$ left-moving zero modes. 

So, in the case $m=0$ or $m=1$ we need $V|_C$ to split trivially
again. In the case $m\geq2$ we need $V|_C$ to be split ``more
trivially'' than the tangent bundle without actually splitting
trivially. We show that there is actually only one possible splitting
type for each $m$.

\begin{prop}
Let $C$ have normal bundle $\O(-1-m)\oplus\O(-1+m)$, with $m>0$. Then
the number of fermion zero modes will force the instanton corrections to
vacuum stability to vanish unless
\begin{equation}
V|_C = \O\oplus\O(1-m)\oplus(-1+m).  \label{eq:msplit}
\end{equation}
\end{prop}

If $m>0$ then $C$ itself contributes to the tangent bundle
deformation count. Instead of (\ref{eq:Tan11}) we now have
\begin{equation}
\xymatrix@C=15mm@R=1mm{
&&\O(1)\ar@{.>}@/_6mm/@<-5mm>[dddd]\\
&&\oplus\\
&&\O(1)\ar@{.>}@/^3mm/@<3mm>[dd]\\
0\ar[r]&\O\ar[r]^-{\left(\begin{smallmatrix}x\\y\\0\\0
\end{smallmatrix}\right)}&
\qquad\quad\oplus\qquad\quad\ar[r]&\cT|_C\ar[r]&0.\\
&&\O(-1-m)\ar@{=>}@/^3mm/@<8mm>[dd]\\
&&\oplus\\
&&\O(-1+m)\\
} \label{eq:Tan12}
\end{equation}
The double arrow shows the Ext's of dimension $m$ that $C$ itself is
producing. Note, however, that this arrow is pointing in the wrong
direction to restrict to anything nonzero on $C$. We show by dotted
arrows the Ext's possibly contributed by intersecting curves that
deform $\cT|_C$. Using the methods of section \ref{sss:defect1} and
the appendix, one can (eventually) show that turning both of these
Ext's on gives (\ref{eq:msplit}).  This is as close to trivial
splitting as we can get, and happens to give exactly the $2m-2$ zero
modes needed above. \QED

\begin{figure}
\begin{center}
\begin{tikzpicture}[x=1.0mm,y=1.0mm]
\path[shape=circle,inner sep=1pt,
    every node/.style={draw,font=\tiny}]
  (0,0) node(a1) {}
  (10,0) node(a2) {}
  (20,0) node(a3) {}
  (10,10) node(a4) {};
\arrja{a2}{a4}{1};
\arrj{a1}{a2}{};
\arrj{a3}{a2}{};
\draw (a1) -- (a4) -- (a3);
\end{tikzpicture}\hspace{15mm}
\begin{tikzpicture}[x=1.0mm,y=1.0mm]
\path[shape=circle,inner sep=1pt,
    every node/.style={draw,font=\tiny}]
  (0,-7) node(a1) {}
  (10,0) node(a2) {}
  (20,-7) node(a3) {}
  (10,10) node(a4) {};
\arrja{a2}{a4}{};
\draw (12,4) node[inner sep=0pt] {$\scriptstyle 2$}; 
\arrj{a1}{a2}{};
\arrj{a3}{a2}{};
\draw (a1) -- (a4) -- (a3);
\end{tikzpicture}\hspace{15mm}
\begin{tikzpicture}[x=1.0mm,y=1.0mm]
\path[shape=circle,inner sep=1pt,
    every node/.style={draw,font=\tiny}]
  (0,-10) node(a1) {}
  (10,0) node(a2) {}
  (20,-10) node(a3) {}
  (10,10) node(a4) {};
\arrja{a2}{a4}{};
\draw (12,3) node[inner sep=0pt] {$\scriptstyle 3$}; 
\arrj{a1}{a2}{};
\arrj{a3}{a2}{};
\draw (a1) -- (a4) -- (a3);
\end{tikzpicture}\hspace{10mm}$\ldots$
\end{center}
\caption{Configurations for non-isolated curves for $m=1,2,3,\ldots$} 
   \label{fig:nonisol}
\end{figure}
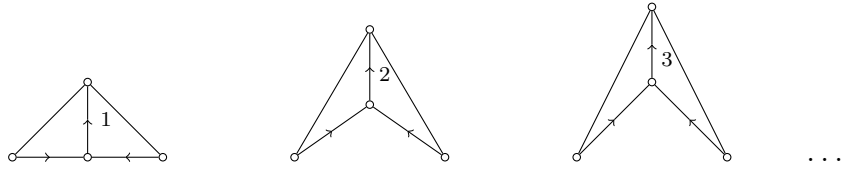

It follows that we get instanton corrections if subdiagrams of the form
in figure \ref{fig:nonisol} appear in our toric diagram. This is
certainly possible in some examples. However, we have the following:
\begin{prop}
The Hilbert scheme resolution of $\C^3/G$ has no worldsheet instanton
corrections to (0,2)-deformations coming from irreducible toric
rational curves.
\end{prop}
To prove this note that it is impossible for any of the diagrams in
figure \ref{fig:nonisol} to appear. This follows from property 1 in
proposition \ref{prop:Craw}. That is, any arrow for the Hilbert scheme
can be followed along a straight line until it reaches an outer
vertex. This is clearly not compatible with figure \ref{fig:nonisol}.

The diagram in figure \ref{fig:min11} is also not possible in the
Hilbert scheme, since dotted lines only come from subdividing
equilateral triangles, and so there are no corrections from
$(-1,-1)$-curves. \QED

Returning to our examples, an examination of figure
\ref{fig:Z11} shows that we never get diagrams of the form in figure
\ref{fig:nonisol} in the $\Z_{11}$ case. Nor
do they occur in the simple defect in figure \ref{fig:defect1}. So
these non-isolated curves do not resolve our discrepancies.

\subsection{Reducible Curves}   \label{ss:reduc}

Rational curves minimize the action in a homotopy class of maps from
$S^2$ into $X$. Unfortunately there is no promise that such a rational
curve is smooth. A nodal rational curve might look like two rational curves
intersecting transversely at a point, i.e., it is reducible. We must
consider such cases as possible instantons.

The problem is that such a rational curve cannot be the image of a
holomorphic map $\pi:\P^1\to X$. To see this note that the inverse
image of one of the components of the target rational curve would need to
be a Zariski closed proper subset of $\P^1$, but the only such subsets
are points. This means the earlier analysis of this paper doesn't really
apply. One possible solution is to consider smooth non-toric curves in
the same family. We cover this in the next section.

The appearance of such nodal curves is very common in the above
examples. For example, in the last diagram in figure \ref{fig:Z11} we
can take the union $C_{38}\cup C_{48}$. Indeed, any time two rational
curves meet at a point, we can take their union.

\subsection{Multiple Covers} \label{ss:mult}

Another possibility is that the worldsheet map $\phi:\Sigma\to C$ is a
multiple cover. Such maps are very important in the $N=(2,2)$ case and
there is no reason to suspect they will not have important effects in the
$(0,2)$ case as well. However, since we are no longer doing
topological field theory, one would not expect such multiple covers to
be given by the same Euler characteristic as in Gromov--Witten theory
\cite{AM:rat}.

Unfortunately, fermion zero mode counting actually suggests more
instanton corrections that we need. Consider a $(-1,-1)$-curve $C$
such that
\begin{equation}
  V|_C = \O\oplus\O(-1)\oplus\O(1). \label{eq:1-1}
\end{equation}
For this embedding, there are 4 right-moving zero modes (from $\cT|_V$)
and 2 left-moving modes. 

If $\phi$ is a double cover, then the degree of each line bundle
doubles. Thus $\phi^{-1}V = \O\oplus\O(-2)\oplus\O(2)$. We now have 8
right-moving and 4 left-moving zero modes. This has a difference of 4
and thus we might na\"\i vely expect instanton corrections.

This is not good. The Hilbert scheme has many tangent bundle
deformations leading to the split (\ref{eq:1-1}) and we don't want to
obstruct these. Since we have merely counted zero modes, we haven't
proven there actually are instanton corrections. Clearly there is more to
be investigated here. In particular, one should try to understand the
integral over the moduli space in analogy with \cite{AM:rat}.

\subsection{Non-toric curves} \label{ss:nontoric}

We can also consider rational curves which are not toric. However, all
rational curves are actually ``seen'' by toric rational curves in the
following sense. There is a $(\C^*)^d$-action on $X$. If a curve is
completely fixed by this action then it is toric. If not, then the
curve must move in a family under this action. We a can then take a
limit of generic one-parameter subfamily of $(\C^*)^d$ to include the
origin. The limit will then be a (not necessarily smooth) toric
rational curve. Thus every rational curve is contained in a family
including a toric rational curve.

A corollary of this statement is that an isolated rational curve must
be toric. That is, all $(-1,-1)$ curves are toric. Thus we have
already studied all instantons given by irreducible isolated rational
curves.

Let us illustrate what can happen with an example in the $\Z_{11}$
resolution appearing last in figure \ref{fig:Z11} (with 45 total deformations). 
The two monomials $x_1x_2^2x_6x_8$ and $x_3^2x_4$ have the same $(\C^*)^5$-charge. 
Thus we may consider the homogeneous ideal
$\cI_{[\alpha,\beta]} = \langle \alpha x_1x_2^2x_6x_8 +\beta
  x_3^2x_4, x_7\rangle$,
for $[\alpha,\beta]$ some point in $\P^1$. The irrelevant ideal for
the case we are considering says that $x_2$, $x_6$, $x_4$ cannot vanish
when $x_7=0$ so the above simplifies to
\begin{equation}
  \cI_{[\alpha,\beta]} = \langle \alpha x_1x_8 +\beta
  x_3^2, x_7\rangle.
\end{equation}
The divisor $x_7=0$ is isomorphic to $\P^2$ with homogeneous
coordinates $[x_1,x_3,x_8]$, so this ideal is the ideal of a plane conic
rational curve, which is smooth if $\alpha$ and $\beta$ are nonzero.

Let $[u,v]$ be homogeneous coordinates on $C$ and put $x_1=\beta u^2$,
$x_3=-\alpha^{\frac12}uv$, $x_8=v^2$ to embed $C$ into $X$. The
restriction of the tangent sheaf to $C$ is now given by
\begin{equation}
\xymatrix@1@C=15mm{0\ar[r]&
\O\ar[r]^-{\left(\begin{smallmatrix}u^2\\uv\\v^2\\0\end{smallmatrix}
\right)}&{\begin{matrix}\O(2)^{\oplus3}\\\oplus\\\O(-6)\end{matrix}}\ar[r]&
\cT|_C\ar[r]&0.}
\end{equation}
This gives $\cT|_C\cong\O(3)\oplus\O(3)\oplus\O(-6)$ and so the adjunction
(\ref{eq:adj}) does not split. Indeed, any plane curve of degree $>1$
will have such a non-splitting \cite{MR0116361}.

In this example, none of the deformations of $\cT$ do anything to this
restriction, so this curve is certainly not a new source of instanton
corrections.

Going to a generic toric limit is easy when given the corresponding
ideal. One simply replaces any non-monomial generator by one of its
monomials. Thus we have $\langle x_1x_8,x_7\rangle$ or $\langle
x_3^2,x_7\rangle$. This corresponds to a reducible rational curve and
a double cover respectively. Thus we see that the the reducible case
and multiple covers are really aspects of the non-toric curve picture.

It is worth pointing out, however, that not every reducible rational
curve is a limit of a family of smooth ones. Consider the union of
$C_{38}$ and $C_{48}$ in the last diagram in figure \ref{fig:Z11}
again. Within the surface $x_8=0$, these curves have self-intersection
$-1$ and $-2$ respectively, and mutual intersection number 1. Thus, their union 
has self-intersection $-1+2-2=-1$. Thus, a smooth curve in the same
family would be a $(-1,-1)$-curve, but this must be isolated which is a
contradiction. Therefore $C_{38}\cup C_{48}$ cannot appear as the toric
limit of a smooth family.

\section{Discussion}  \label{s:conc}

We have had modest success in considering the effects of smooth
toric rational curves. In the case of Hilbert schemes, where we do not
need any instanton corrections, we do not get any. Furthermore, in
examples we considered where we needed instanton corrections we
frequently got some. The only problem is that the instantons we
analyzed did not reduce the moduli space by enough dimensions to
agree with conformal field theory.

Clearly there is a need to understand instantons on reducible rational
curves. We suspect that these are the source of the discrepancies we
find. Another direction which should be pursued is an analysis of the
integration over moduli space required to understand multiple covers. 

\section*{Acknowledgments}

We thank R.~Plesser for useful discussions. This work
was partially supported by NSF grants DMS--0905923 and DMS--1207708.
Any opinions, findings, and conclusions or recommendations expressed
in this material are those of the authors and do not necessarily
reflect the views of the National Science Foundation.

\section*{Appendix: Extensions on $\P^1$}   \label{s:ext}

In this appendix we review extensions of line bundles on $\P^1$, which
are central to this paper. Let $\cE$ satisfy
\begin{equation}
\xymatrix@1{0\ar[r]&\O(-n)\ar[r]&\cE\ar[r]&\O(n)\ar[r]&0.}
\end{equation}
Then what are the allowed sheaves $\cE$? The answer is
\begin{equation}
\O(-n)\oplus\O(n),\:\hbox{or}\:\O(-n+1)\oplus\O(n-1),
\ldots,\:\hbox{or}\:\O\oplus\O.  \label{eq:elist}
\end{equation}
The first in this list is the trivial extension. In the space of
possible extensions, we want to show that these extensions get more
and more likely as we go down the list in the sense that they occupy a
higher-dimensional set. The generic extension will then be the ``trivial'' split
$\O\oplus\O$.

The space of extensions is given by
\begin{equation}
  \Ext^1(\O(n),\O(-n)) = H^1(\P^1,\O(-2n)) = \C^{2n-1},
\end{equation}
but we need to find exactly which extensions elements of this
extension group give. We do this in the language of the derived
category (or one could use push-out diagrams as in page 77 of
\cite{Wei:hom}.) The result is easy to state. If
$f:\O(n)[-1]\to\O(-n)$ represents an element of
$\Ext^1(\O(n),\O(-n))$, then the desired extension $\cE$ is simply the
mapping cone of $f$. The only work required is to explicitly relate
this mapping cone to the possibilities listed in (\ref{eq:elist}).
Rather than doing this in generality we analyze the case $n=2$. The
same methods can be employed for any $n$.

Since
\begin{equation}
\xymatrix@1@C=12mm{0\ar[r]&\O\ar[r]^-{\left(\begin{smallmatrix}x\\y
\end{smallmatrix}\right)}&
\O(1)^{\oplus2}\ar[r]^-{\left(\begin{smallmatrix}-y&x\end{smallmatrix}\right)}&
\O(2)\ar[r]&0,}  \label{eq:O2}
\end{equation}
we have an isomorphism in $\DC(\P^1)$ between $\O(2)$ and the complex
given by the two left terms above. We may now apply $\otimes\O(-1)$ to
(\ref{eq:O2}) to get an isomorphism between $\O(1)$ and the complex
$\O(-1)\to\O^{\oplus2}$. Using this to eliminate $\O(1)$ and then using
Gaussian elimination to find a
minimal presentation we get a short exact
sequence
\begin{equation}
\xymatrix@1@C=12mm{0\ar[r]&
\O(-1)^{\oplus2}\ar[r]^-{\left(\begin{smallmatrix}x&0\\y&x\\0&y\end{smallmatrix}
\right)}&\O^{\oplus3}\ar[r]&\O(2)\ar[r]&0.}
\end{equation}
Repeating this process we get
\begin{equation}
\xymatrix@1@C=12mm{0\ar[r]&
\O(-2)^{\oplus3}\ar[r]^-{\left(\begin{smallmatrix}x&0&0\\y&x&0\\0&y&x\\
0&0&y\end{smallmatrix}
\right)}&\O(-1)^{\oplus4}\ar[r]&\O(2)\ar[r]&0.}
\end{equation}
This isomorphism between $\O(2)$ and the complex given by the two left
terms above allows us to explicitly see the required map $f$. The
result is that $\cE$ is given by
\begin{equation}
\xymatrix@1@C=15mm{0\ar[r]&\O(-2)^{\oplus3}
\ar[r]^-{\left(\begin{smallmatrix}a&b&c\\x&0&0\\y&x&0\\0&y&x\\
0&0&y\end{smallmatrix}
\right)}&{\begin{matrix}\O(-2)\\\oplus\\\O(-1)^{\oplus4}\end{matrix}}
\ar[r]&\cE\ar[r]&0,}  \label{eq:E2}
\end{equation}
where $(a,b,c)\in\Ext^1(\O(2),\O(-2))\cong \C^3$.

Now to determine which possibility we have, note that we can
distinguish between them by computing $H^0(\cE(-1))$. This, in turn we
get from the long exact sequence of (\ref{eq:E2}). We get the explicit
map between $H^1(\O(-3)^{\oplus3})$ and
$H^1(\O(-3)\oplus\O(-2)^{\oplus4})$ 
using monomials $x^\alpha y^\beta$, with $\alpha,\beta<0$ to represent
local cohomology. The result is that $H^0(\cE(-1))$ is given by the
kernel of the matrix
\begin{equation}
A=\begin{pmatrix}
a&0&b&0&c&0\\
0&a&0&b&0&c\\
1&0&0&0&0&0\\
0&1&1&0&0&0\\
0&0&0&1&1&0\\
0&0&0&0&0&1
\end{pmatrix}.
\end{equation}
$A$ has rank 4 if and only if $a=b=c=0$, and determinant $b^2-ac$. Thus
we get the final result
\begin{equation}
\cE = \begin{cases}
\O(-2)\oplus\O(2)\qquad&\hbox{if}\quad a=b=c=0,\\
\O(-1)\oplus\O(1)\qquad&\hbox{if otherwise and}\quad b^2-ac=0,\\
\O\oplus\O\qquad&\hbox{otherwise}.
\end{cases}
\end{equation}
As promised, $\O\oplus\O$ occurs generically. Furthermore
$\O(-1)\oplus\O(1)$ happens in a codimension one subspace of
$\Ext^1(\O(2),\O(-2))$ and $\O(-2)\oplus\O(2)$ only occurs in
codimension 3.


\end{document}